\documentclass[final,5p,times,twocolumn,sort&compress]{elsarticle}

\usepackage{graphicx}
\usepackage{epsfig}
\usepackage{amssymb}
\usepackage{amsmath}
\usepackage{slashed}
\usepackage{units}
\usepackage{textcomp}
\usepackage{hyperref}
\usepackage{array}
\usepackage{lmodern}
\usepackage{url}
\usepackage{lineno}
\usepackage{tabularx}
\usepackage{threeparttable}
\usepackage{booktabs}
\usepackage{dcolumn}

\biboptions{sort&compress}

\journal{Physics Letters B}

\begin{document}

\begin{frontmatter}

\title{A New Measurement of the Intruder Configuration in $^{12}$Be}

\author[pku]{J. Chen}
\author[pku]{J. L. Lou\corref{cor}}%
 \ead{jllou@pku.edu.cn}

\author[pku]{Y. L. Ye}%
%\ead{yeyl@pku.edu.cn}

\author[pku]{Z. H. Li}%
\author[bh]{D. Y. Pang}%
\author[sz]{C. X. Yuan}
\author[pku]{Y. C. Ge}%
\author[pku]{Q. T. Li}%
\author[pku]{H. Hua}%
\author[pku]{D. X. Jiang}%
\author[pku]{X. F. Yang}%
\author[pku]{F. R. Xu}%
\author[pku]{J. C. Pei}%
\author[pku]{J. Li}%
\author[pku]{W. Jiang}%
\author[pku]{Y. L. Sun}%
\author[pku]{H. L. Zang}%
\author[pku]{Y. Zhang}%
\author[rcnp]{N. Aoi}%
\author[rcnp]{E. Ideguchi}%
\author[rcnp]{H. J. Ong}%

\author[riken]{J. Lee}%
\author[riken]{J. Wu}%
\author[riken]{H. N. Liu}%
\author[riken]{C. Wen}%
\author[rcnp]{Y. Ayyad}%
\author[rcnp]{K. Hatanaka}%
\author[rcnp]{T. D. Tran}%
\author[rcnp]{T. Yamamoto}%
\author[rcnp]{M. Tanaka}%
\author[rcnp]{T. Suzuki}%

\cortext[cor]{Corresponding author}

\address[pku]{%
 School of Physics and State Key Laboratory of Nuclear Physics and Technology, Peking University, Beijing 100871, China}%
\address[bh]{%
School of Physics and Nuclear Energy Engineering, Beijing Key Laboratory of Advanced Nuclear Materials and Physics,  Beihang University, Beijing 100191, China}%
\address[sz]{
Sino-French Institute of Nuclear Engineering and Technology, Sun Yat-Sen University, Zhuhai 519082, China}
\address[rcnp]{%
Research Centre for Nuclear Physics, Osaka University, Osaka 567-0047, Japan}%
\address[riken]{%
 RIKEN (Institute of Physical and Chemical Research), 2-1 Hirosawa, Wako, Saitama 351-0198, Japan}%


\begin{abstract}
A new $^{11}$Be($d,p$)$^{12}$Be transfer reaction experiment was carried out in inverse kinematics at 26.9$A$ MeV, with special efforts devoted
 to the determination of the deuteron target thickness and of the required optical potentials from the present elastic scattering data. In addition
 a direct measurement of the cross section for the 0$_2^+$ state was realized by applying an isomer-tagging technique. The $s$-wave spectroscopic
  factors of 0.20$^{+0.03}_{-0.04}$ and 0.41$^{+0.11}_{-0.11}$ were extracted for the 0$_1^+$ and 0$_2^+$ states, respectively, in $^{12}$Be.  Using the ratio of these
  spectroscopic factors, together with the previously reported results for the $p$-wave components, the single-particle component intensities in
  the bound 0$^+$ states of $^{12}$Be were deduced, allowing a direct comparison with the theoretical predictions. It is evidenced that the
   ground-state configuration of $^{12}$Be is dominated by the $d$-wave intruder, exhibiting a dramatic evolution of the intruding mechanism
    from $^{11}$Be to $^{12}$Be, with a persistence of the $N = 8$ magic number broken.
\end{abstract}


\end{frontmatter}

\section{Introduction}
According to the well-established mean field framework for nuclear structure, nucleons (protons or neutrons) are filling in the single-particle
orbitals grouped into shells characterized by the conventional magic numbers \cite{Kanungo-2013}. However, for nuclei far from the $\beta$-stability
line,  especially those in the region of light nuclei where the concept of a mean field is less robust, the exotic rearrangement of the single-particle
configuration often appears and may result in vanishing or changing of the magic numbers. One widely-noted example is the ground state (g.s.) of
the one-neutron-halo nucleus $^{11}$Be, which possesses an unusual spin-parity of $\rm{1/2}^+$, being dominated ($\sim71\%$) by an intruding
$\rm{1s}_{1/2}$ neutron coupled to a $^{10}$Be$(0^+)$ core \cite{Schmitt,Aumann}. Obviously the prominent appearance of the $s$-wave in the
 g.s. of $^{11}$Be is responsible for the formation of its novel halo structure \cite{Tanihata}.

The immediate question goes into the single-particle configuration of $^{12}$Be, having one more valence neutron outside the $^{10}$Be core.
This neutron-rich nucleus has four particle-bound states, namely the g.s.($0^+$), and the excited states at 2.107~($2^+$), 2.251~($0^+$)
 and 2.710 MeV~($1^-$) \cite{Shimoura-2007}. The relatively low energies of the latter three states imply the breakdown of the  $N = 8$ magic number and the strong intruder from the upper $sd$-shell \cite{Iwasaki-1,Iwasaki-2,Shimoura-2003,Shimoura-2007,Imai-2009}, leading to the growth of
 other non-shell-like structure in this nucleus \cite{Yang,Yang2015}. Since Barker's early work in describing the isospin $T = 2$ states
 of the mass $A = 12$ nuclei with a mixed configuration \cite{Barker-1976}, substantial theoretical studies have been devoted to the
 spectroscopic studies of the low-lying states in $^{12}$Be. To date most studies agree on the large probability ($\ 60\%$) of intruder
 from the $sd$-shell, but the relative importance of the $s$- and $d$-components remains a subject of active investigation \cite{Kanungo-2013}.
  A standard way to describe the intruding effects around $N = 8$ is to use the configuration mixing $\alpha(s^2) + \beta(d^2) + \gamma(p^2)$,
  with $\alpha$, $\beta$ and $\gamma$ the normalized intensities (percentages) of the respective components for valence neutrons in 0$^+$-states
  outside the $^{10}$Be core \cite{Barker2009,Fortune2009}. In principal there should be three 0$^+$ states in this $p-sd$ model space, but only
   the lowest two have been found in the bound region.  The third 0$^+$ state was predicted to appear in a wide energy range of
   3$\sim$9 MeV \cite{Barker-1976,Fortune2006,Barker2009,Smith}, but to date it has not been identified experimentally. Therefore
   in the present work we focus on the lowest two $0^+$ states only. Table~\ref{tab:table1} (upper panel) summarizes the individual
    intensities from the shell model calculations by Barker  \cite{Barker-1976} and Fortune $et\ al.$ \cite{Fortune2006}, the three-body
     model predictions by Nunes $et\ al.$ \cite{Nune2002} and Redondo $et\ al.$ \cite{Romero-Redondo}, the nuclear field theory approach
     by Gori $et\ al.$ \cite{Gori2004}, and the random-phase approximation by Blanchon $et\ al.$  \cite{Blanchon}. The results are quite
     disparate in terms of the dominant component of each state. For instance the $s$-wave intensity in the $0_1^+$ g.s. ranges from $23\%$
     up to 76$\%$, resulting in active disputing \cite{Barker2009,Fortune2009}. In fact the model calculation of the configuration admixture
      depends on various basic physics ingredients, such as the particle-separation energy, the deformation of the nucleus, the core-nucleon
       potential and wave functions, the effective pair interaction, the interplay between the collective motion and the valence nucleons,
       and so on\cite{Kanungo-2013,Gori2004}. Particularly the ratio of $s^2$ to $d^2$ is sensitively regulated by the core-nucleon Hamiltonian
        and the nucleon-nucleon residual interaction \cite{Sherr1999}.

\begin{table}
 \caption{\label{tab:table1}  Intensities of the $s$($\alpha$)-, $d$($\beta$)- and $p$($\gamma$)-components in the first two $0^+$ states
 of $^{12}$Be, predicted by various model calculations with the same normalization scheme (upper panel). The selected experimental results
  are also presented (lower panel), as explained in the text. }
%  The error bars are omitted here for simplicity.}
%\begin{ruledtabular}
  \begin{threeparttable}
  \begin{tabular}{p{0.8cm}p{0.8cm}p{0.8cm}p{0.8cm}p{0.8cm}p{0.8cm}p{1.3cm}}
  \toprule
  \multicolumn{3}{c}{0$_1^+$}&\multicolumn{3}{c}{0$_2^+$} &\\
  $\alpha_1(\%)$&$\beta_1(\%)$& $\gamma_1(\%)$ &
  $\alpha_2(\%)$ &$\beta_2(\%)$&$\gamma_2(\%)$ &  Ref \\
  \hline
 33 & 29 & 38 & 67 & 10 & 23 &\cite{Barker-1976} \\
 53 & 15 & 32 & 25 & 7 & 68 &\cite{Fortune2006,Barker2009}\\
 31 & 42 & 27 &    &    &    &\cite{Nune2002,Pain} \\
 67$\sim$76 & 10$\sim$13 & 13$\sim$19 & 15$\sim$23 & 6$\sim$8 & 71$\sim$78 &\cite{Romero-Redondo} \\
  23 & 48 & 29 &    &    &    &\cite{Gori2004} \\
 25$^a$ & 21$^a$ & 54$^a$ & 62$^a$ & 0$^a$ & 38$^a$ &\cite{Blanchon}\\

 \hline
% 33$^b$ & 38$^b$ & 29$^b$ &      &      &      &\cite{Navin,Pain}\\
  33$^b$ & 38$^b$ & 29$^b$ &      &      &      &\cite{Navin,Pain}\\
        &        & 24$\pm$5$^c$     &      &      & 59$\pm$5$^c$   &\cite{Meharchand}\\
% 16     &        &        & 32   &      &      & this work \\
 19$\pm$7        &   57$\pm$7       &        & 39$\pm$2   &   2$\pm$2    &      &this work \\
\bottomrule
\end{tabular}
 \begin{tablenotes}
        \footnotesize
       \item{a} from Table.2 and Table.3 of Ref. \cite{Blanchon}.
       \item{b} using SFs of 0.42, 0.48 and 0.37 for $s$-, $d$- and $p$-components \cite{Navin,Pain,Kanungo}, respectively,
       which are normalized to their sum to give the intensities \cite{Barker2009}.
       \item{c} $p$-wave intensities extracted from a charge-exchange experiment \cite{Meharchand}.
 \end{tablenotes}
  \end{threeparttable}
%\end{ruledtabular}
\end{table}

As discussed in detail in Refs.\cite{Barker2009,Fortune2009,Fortune2012,Kanungo-2013}, various experiments have been carried
out to quantify the intruder strengths. Here in Table \ref{tab:table1} (lower panel) are listed only those sensitive to individual
 structure component.
One-neutron knockout reactions were performed for $^{12}$Be to extract spectroscopic factor (SF) of each single-particle orbit
\cite{Navin,Pain}. The comparison to the theoretical intensities can be made by normalizing to the sum of the three SFs, similar
 to the way used in row N of Table I in Ref.\cite{Barker2009}. The obtained values show almost equivalent intensities for the $s$-,
  $d$- and $p$-orbital in the g.s. of $^{12}$Be. It was noticed that the $^{12}$Be beam used in the knockout reaction may be
   in both the g.s. and the long-lived isomeric $0_2^+$ state, leading to a reduced strength difference between the two $0^+$
   states \cite{Kanungo}. One-neutron transfer reaction, namely $^{11}$Be($d,p$)$^{12}$Be at 5$A$ MeV, was carried out to populate
    the $s$-component in the first two $0^+$ states of $^{12}$Be. The obtained SFs are 0.28$^{+0.03}_{-0.07}$ and 0.73$^{+0.27}_{-0.40}$,
     respectively. This experiment was later on questioned for the possible contamination of the $\rm (CD_2)$$_n$ target and the large
     uncertainties in extracting SFs from the undistinguishable 0$_2^+$ and $2^+$ states \cite{Fortune2012}. Another one-neutron transfer
      experiment at 2.8$A$ MeV was then performed with a clear separation of all low-lying excited states by incorporating the $\gamma$-ray
      detection \cite{Johansen}. The extracted SFs (set III) are $0.15_{-0.05}^{+0.03}$ and $0.40_{-0.09}^{+0.13}$, respectively, for two
       low-lying $0^+$ states. This experiment suffered from a very low beam energy, leading to an effective detection outside the most
        sensitive angular range, especially for the $0_2^+$ state. Due to the lack of proper normalization procedures for these transfer
         reactions, it would be difficult to compare their SF results with other measurements or to each other \cite{Kay-2013}.
Recently the $p$-wave intensities for the two low-lying $0^+$ states were determined from a charge-exchange experiment \cite{Meharchand},
which are listed also in Table \ref{tab:table1}.
It is evident that more measurements are urgently needed to clarify the theoretical deviations and the experimental
 ambiguities \cite{Fortune2012,Kanungo-2013}. In this letter, we report on a new measurement of the $^{11}$Be($d,p$)$^{12}$Be transfer
  reaction, with special measures taken to deal with the questioned experimental uncertainties.

\section{Experimental setup}
\begin{figure}
  \includegraphics[width=0.45\textwidth]{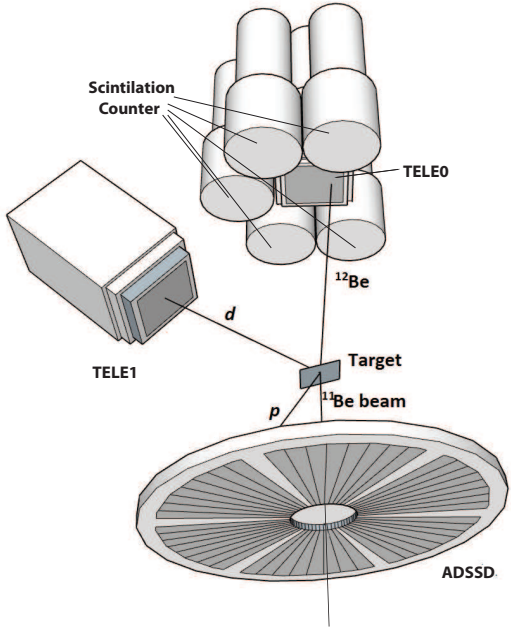}\\
  \caption{Schematic view of the experimental setup (more details in Ref.\cite{Chen-1}).}
    \label{setup}
\end{figure}

\begin{figure}
  \includegraphics[width=0.50\textwidth]{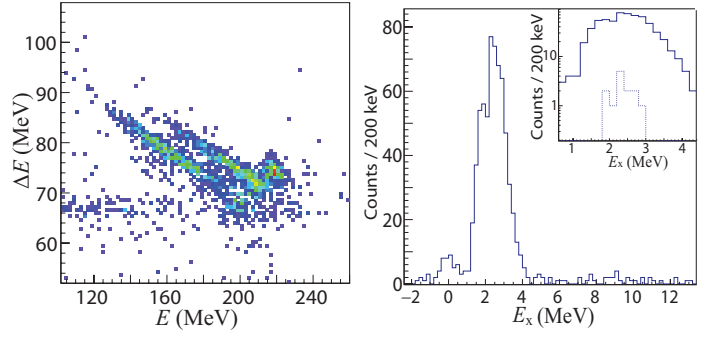}\\
  \caption{(a) PID spectrum taken by the TELE0, using energy loss $\Delta E$ in the DSSD versus remaining energy in the SSD, in
  coincidence with protons recorded by the ADSSD. (b) The excitation energy spectrum for bound states in $^{12}\rm Be$, deduced from
  recoil protons in coincidence with $^{12}\rm Be$ isotope in the TELE0 (solid curves). The dotted curve in the inset shows the events
  having the further coincidence with the 0.511 MeV $\gamma$-rays detected by the scintillation counters around the TELE0.}
    \label{PID}
\end{figure}

 The experiment was carried out at the EN-course beam line, RCNP (Research Center for Nuclear Physics), Osaka University~\cite{Shimoda}.
 A $^{11}\rm Be$ secondary beam at 26.9$A$ MeV with an intensity of $10^4$ particles per second (pps) and a purity of about 95$\%$ was
  produced from a $^{13}$C primary beam impinging on a Be production target with a thickness of 456 mg/cm$^2$. The energy of the secondary
  beam was chosen considering the effective detection of the recoil protons at backward angles, the availability of the primary beam, and the
   validation of the transfer reaction mechanism. A schematic view of the detection system is shown in Fig.\ref{setup} (with more details in
   Ref.\cite{Chen-1}). Elastic scattering of $^{11}$Be from protons or deuterons was measured by using a $\rm (CH_2)$$_n$ (4.00 $\rm mg/cm^2$)
   or a $\rm (CD_2)$$_n$ (4.00 $\rm mg/cm^2$) target, respectively, with the background subtraction provided by C-target runs
   \cite{Chen-1,Chen-2}. The inevitable hydrogen contamination in the $\rm (CD_2)$$_n$ target was found to be $9.5\pm0.6 \%$ out
   of the total deuterium contents, determined by the number of recoil protons relative to those from the known $\rm (CH_2)$$_n$
   target \cite{Chen-2}. The incident angle and the hit position on the target were determined by two parallel-plate avalanche counters
   (PPAC) placed upstream of the target (not shown in the figure), with resolutions (FWHM) less than 0.3$^\circ$ and 2.0 mm, respectively.
   The backward emitted protons were detected using a set of the annular double-sided silicon-strip detector (ADSSD in Fig.\ref{setup})
   composed of six sectors, each divided into sixteen 6.4-mm-wide rings on one side and 8 wedge-shaped regions on the other side. This
   annular detector has an inner and an outer radii of 32.5 mm and 135 mm, respectively, covering laboratory angles
   of $165^{\circ}\sim 135^{\circ}$ relative to the beam direction. The energy detection threshold was set at 1.0 MeV,
   allowing to cut off the noise while retaining a high sensitivity for protons related to interested excited states in $^{12}$Be.
    The ADSSD provided also good timing signals with a resolution ($\sim 2$ ns) good enough to reject protons not coming from the
    target.  The forward moving projectile-like fragments were detected and identified by a set of charged-particle telescope
    (TELE0 in Fig.\ref{setup}) composed of a double sided silicon-strip detector (DSSD) of 1000 $\mu$m thick and two layers of
    large size silicon detector (SSD), each having a thickness of 1500 $\mu$m. This telescope has an active area of
    $62.5 \times 62.5$ mm$^2$ ($32 \times 32$ strips) and was centered at the beam direction ($0^0$) at a distance of 200 mm
     down stream from the target. A particle identification (PID) spectrum, taken by the TELE0 and in coincidence with protons
      in the ADSSD, is shown in Fig.~\ref{PID}(a). $^{12}$Be in the figure must come from the ($d, p$) transfer reaction,
       whereas $^{11}$Be and $^{10}$Be, with much broader energy spread, are most likely related to the neutron decay following
       the population of unbound states in $^{12}$Be. The coincidence with the backward-emission protons is essential here to
        avoid the large background arising from the direct beam \cite{Pain}.

Gated on $^{12}$Be in TELE0, protons are the only charged particles being detected in the ADSSD and therefore their kinematics can
 be mapped out based on the detected energies and angles. The excitation energy in $^{12}$Be can then be deduced from the recoil protons,
  as shown in Fig.\ref{PID}(b). Monte Carlo simulations were conducted to estimate the resolution and efficiency as a function of the
  excitation energy. An integrated energy resolution of about 1.1 MeV(FWHM) is in agreement with the width of the g.s. peak (centered
  at 0.0 MeV) in Fig.\ref{PID}(b). Although the number of counts in this peak is relatively small, its significance is clear due to
   the very low background. A large and broad peak stands between 1 and 4 MeV, contributed from the unsolved three states at 2.107, 2.251
    and 2.710 MeV in $^{12}$Be. It is worth noting that protons belonging to the g.s. peak in Fig.\ref{PID}(b) have higher energies in
     the ADSSD detector and thus are almost free from the detection loss. Also background counts were checked by employing a carbon target.

\begin{figure}
%   Requires \usepackage{graphicx}
  \includegraphics[width=0.43\textwidth]{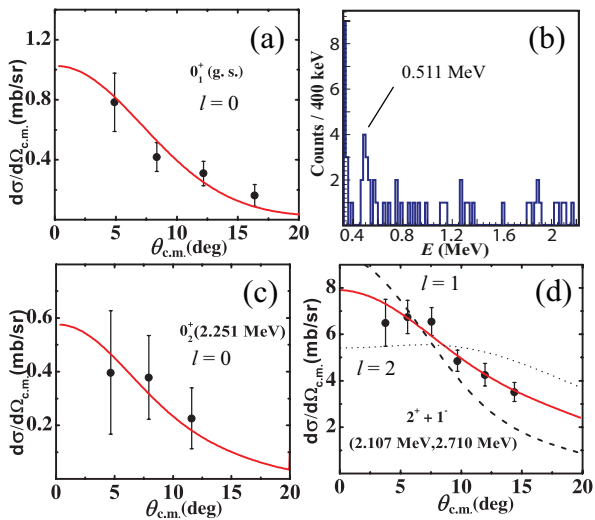}\\
  \caption{ Measured differential cross sections of the $^{11}\rm Be$$(d,p)$$^{12}\rm Be$ reaction at 26.9$A$ MeV (solid dots), together
  with the FR-ADWA
  calculations (curves as described in the text), for (a) the g.s. ($0_1^+$), (c) the isomeric state ($0_2^+$),  and (d) the summed $2^+$
   and $1^-$ states. $l$ in (a), (c) and (d) denotes the transferred orbital angular momentum into the final state of $^{12}$Be. (b) is
   dedicated to the $\gamma$-ray energy spectrum
  in coincidence with $^{12}$Be + $p$ events. }\label{DCS}
\end{figure}

 A special isomer-tagging method was used to discriminate the $0_2^+$ state from the broad excitation-energy peak (Fig.\ref{PID}(b)).
 The method relies on its well-known isomeric property: a life-time of 331 $\pm$ 12 ns \cite{Shimoura-2007} and an E0-decay (via e$^+$e$^-$
 pair emission) branching ratio of 83 $\pm$ 2 $\%$ \cite{Shimoura-2003}. $^{12}$Be($0_2^+$) isomers were stopped in the TELE0 and the subsequently
 emitting $\gamma$-rays, particularly the 0.511 MeV ones from the e$^+$-annihilations, were measured by an array of six large-size NaI(Tl)
 scintillation detectors surrounding or at the back of the TELE0 (Fig.\ref{setup}). This kind of  decay-tagging method has been successfully
 applied in many particle-emission experiments \cite{Paul-1995,Lou-2007,Li-2009}.  The $^{12}$Be + $p$ + $\gamma$ triple-coincidence was
 realized based on the good timing signals generated from the strips in the TELE0 and the ADSSD, and from the scintillation detectors,
  respectively. A time window of 3 $\rm{\mu s}$ for the triple-coincidence was applied, which covers about 9 times of the decay
  half-live (331 ns) of the $0_2^+$ state. The $\gamma$-energy spectrum of these triple-coincidence events is presented in Fig.\ref{DCS}(b),
  with the 0.511 MeV $\gamma$-ray peak (between 0.4 and 0.6 MeV) standing well above the background. The time distribution of these 0.511 MeV
  $\gamma$-rays follows approximately the exponential-decay curve with an extracted half-life of 270 $\pm$ 120 ns, being consistent with the
   reported value \cite{Shimoura-2007} within the error bar. The source of these coincidentally observed 0.511 MeV $\gamma$-rays were checked
   against all possible contaminations, such as the random or accidental coincidences, target impurities, event mixing and so on. These can
   be realized by selecting various event samples, other than the targeted one, to build the similar coincidences. Furthermore, the
   possible 0.511 MeV $\gamma$-rays cascaded, or indirectly produced, from other decay-chains in $^{12}$Be were analyzed by realistic
   Monte Carlo simulations. It turned out that all these backgrounds are negligible, mostly attributed to the strict triple-coincidence
   $^{12}$Be + $p$ + $\gamma$ and the detector-setup scheme. Of course this strict coincidence would lead to a reduction of the detection efficiency
    and hence the event statistics. However the observation of the $0_2^+$ isomer-decay is still at very high significance (at least
    3.4$\sigma$, or $>99.9\%$ confidence level), owing to the very low background. The triple-coincidence events are presented in the
    insert of Fig.\ref{PID}(b)(dotted curve), which do concentrate around the excitation energy of the $0_2^+$ state (2.251 MeV).

The detection efficiency in the present work for the 0.511 MeV $\gamma$-rays, produced from the positron annihilation, is determined to be
23 $\pm$ 1 $\%$, based on realistic Monte Carlo simulations using the GEANT4 code \cite{Agostinelli}.

\section{Experimental result}
Differential cross sections for the $^{11}$Be($d,p$)$^{12}$Be transfer reaction at 26.9$A$ MeV are presented in Fig.~\ref{DCS}, deduced
from the recoil protons and gated on the excited state in $^{12}$Be. The g.s. events are selected by a cut from -1.0 to 0.6 MeV on the
excitation energy spectrum (Fig.\ref{PID}(b)). A gate between 0.4 and 0.6 MeV on the $\gamma$-ray energy spectrum (Fig.~\ref{DCS}(b)) is
applied to select the isomeric $0_2^+$ state. $2^+$ and $1^-$ states are still indistinguishable from the excitation energy spectrum
 (Fig.~\ref{PID}(b)) and the summed cross sections are plotted in
Fig.~\ref{DCS}(d) with those for $0_2^+$ state subtracted. The error bars in the figure are statistical only. The systematic error is
less than $10\%$, taking into consideration the uncertainties in the detection efficiency determination($\sim 5\%$), the (CD$_2$)$_n$
 target thickness ($\sim 2\%$), and the cuts on the PID spectrum ($\sim 4\%$) and on the excitation energy spectrum ($\sim5\%$).

To extract the SFs, theoretical calculations were performed by using the code FRESCO~\cite{fresco}, which incorporates approaches such
as the distorted wave Born approximation (DWBA) or the finite-range adiabatic distorted wave approximation (FR-ADWA).
Due to the uncertainties in DWBA calculation associated with the applied optical potentials (OPs) \cite{Schmitt,Johansen,Winfield}, we
adopt the FR-ADWA method, which uses nucleonic potentials, includes explicitly the deuteron breakup process and can provide consistent
 results for ($d$, $p$) transfer reactions \cite{Schmitt}. In the present work the \emph{p} + \emph{n} potential is given by the Reid
 soft-core interaction \cite{Reid}.
A Woods-Saxon form was used for the $^{11}$Be + $n$ binding potential, with a fixed radius and diffuseness of 1.25 fm and 0.65 fm,
respectively. These geometrical parameters were widely adopted for loosely-bound states in light nuclei
\cite{Schmitt,20o,Margerin-26Al,Tsang-excited}. The well depth of this binding potential was adjusted to reproduce the correct
 excitation energies \cite{Kanungo}, and the obtained values are 65.18 MeV and 56.49 MeV, respectively, for the 0$_1^+$ and $0_2^+$ states.
  The entrance channel OP is obtained by folding the $^{11}$Be + $p$ and $^{11}$Be + $n$ potentials, with the former extracted from
  the present elastic-scattering data \cite{Chen-1} and the latter from global potentials \cite{CH89,KD02}. As a matter of fact the
  currently extracted potential is just the global one (CH89) but with two normalization factors, namely 0.78 and 1.02, applied to the depths
  of the real and imaginary parts of the potential, respectively. These normalization factors are necessary for weakly-bound nuclei and the
  currently adopted factors are close to the averaged ones in the literature \cite{Chen-1}. The exit channel OP is extracted from the data
  reported in Ref.\cite{Korsheninnikov} by using the same method as for the $^{11}$Be + $p$ elastic-scattering data.

The results of FR-ADWA calculations, multiplied by the SFs for the selected single-particle component, are fitted to the experimental
 data by the standard $\chi^2$ minimization method \cite{Kanungo}, and the results are shown in Fig.~\ref{DCS}.
Data in Fig.~\ref{DCS}(d) for the mixed $2^+$ and $1^-$ states are fitted by the weighted sum
of $ S1 \cdot (^{11}$Be $\otimes n(1d_{5/2}) ) + S2 \cdot (^{11}$Be $\otimes n(1p_{1/2}))$, where $S1$ and $S2$ are SFs for the
$d$-wave and $p$-wave neutrons in the low-lying $2^+$ and $1^-$ states in $^{12}$Be, respectively.
The best fit (red solid curve in Fig.\ref{DCS}d) is obtained by $S1$ = 0.26 $\pm$ 0.05 and $S2 $ = 0.76 $\pm$ 0.17, with the error bars
corresponding to
a 68.3$\%$ confidence level \cite{Kanungo}. If only one component was used, the result is represented by the dotted or dashed curve for
a pure $d$-wave with SF = 0.5 or a pure $p$-wave with SF = 1.4, respectively. We notice that the $2^+$ state was resolved in an previous
measurement \cite{Johansen}, but the unfavorable angular coverage of the data did not allow a unique extraction of the SF. Our
SF result for the $d$-wave component in the $2^+$ state, 0.26 $\pm$ 0.05, is consistent with two out of four sets of results reported in Ref.\cite{Johansen} for various selections of optical potentials, namely 0.30 $\pm$ 0.10 (set II), and 0.40 $\pm$ 0.10 (set III).

 The extracted $s$-wave SFs for the 0$_1^+$ and 0$_2^+$ states are 0.20 $^{+0.03}_{-0.04}$ and 0.41 $^{+0.11}_{-0.11}$, respectively,
 with the error bars corresponding to a 68.3$\%$ confidence level \cite{Kanungo}.  These results are compatible with those obtained from the previous transfer experiments within the error bars  \cite{Kanungo,Johansen}, although the normalization of the SFs for each measurement was not obtained. Since we have resolved the $0_2^+$ state by using the implantation-decay technique and applied the more suitable FR-ADWA analysis \cite{Schmitt}, the currently extracted SF should be more reliable. It should be worth noting that, although the cross section for the 0$_2^+$ state looks smaller than that for the 0$_1^+$ state, its SF is two times as big as that of the latter one. This is essentially attributed to the
 large reduction of the calculated cross sections for the halo-like states. This behavior was also clearly exhibited in the
 similar reaction $^{15}$C($d$, $p$), in which the $s$-wave SFs of 0.60 $\pm$ 0.13 and 1.40 $\pm$ 0.31 were extracted for the
 first and second 0$^+$ states in $^{16}$C \cite{16C-Wuosmaa}. This difference in cross sections for various final states may
 depend also on the incident energy \cite{CRC-8He} due naturally to the match of the internal and external waves. However,
 since this energy dependence happens for both the measurement and the proper calculation, the SFs, at least for its relative
 or normalized values, should be stable within a relevant energy range\cite{CRC-8He}.

In order to compare our SF results with those from theoretical calculations and from other measurements, the conversion into relative
intensities (percentages) is required \cite{Kay-2013}. Since the necessary quantities related to the sum rule were not measured, we
rely on the ratio of SFs for the $0_1^+$ and $0_2^+$ states, which is independent of the normalization factors. Using the standard
 method proposed by Barker \cite{Barker-1976}, the wave functions of the two low-lying $0^+$ states can be
 written as $|0_i^+\rangle = a_i|1s^2_{1/2}\rangle + b_i|0d^2_{5/2}\rangle + c_i|0p^2_{1/2}\rangle$ ($i = 1,2$), with the
 normalization relations $a_i^2 + b_i^2 + c_i^2 = \alpha_i + \beta_i + \gamma_i = 1$ and the orthogonal requirement
  $a_1 \ast a_2 + b_1 \ast b_2 + c_1 \ast c_2 = 0$. From the present measurement we
  have $\alpha_1 / \alpha_2 = 0.20/0.41 = 0.49^{+0.15}_{-0.16}$. The errors are statistic only. The systematic uncertainty of this ratio is estimated to be less than $\pm$ 7\%, due basically to the possible choices of the optical potentials. Previously the $0p_{1/2}$-wave strengths in the two low-lying $0^+$ states
  of $^{12}$Be were investigated via a charge-exchange reaction experiment $\rm{^{12}B(^7Li,^7Be)^{12}Be}$ \cite{Meharchand}.
  The extracted values are $\gamma_1 = 0.24$ and  $\gamma_2 = 0.59$ within the $p$-$sd$ model space. Combining all these conditions,
  the intensities in the above normalization equations can be deduced: $\alpha_1 = 0.19 \pm 0.07$, $\beta_1 = 0.57 \pm 0.07$, $\gamma_1 = 0.24 \pm 0.05$, $\alpha_2 = 0.39 \pm 0.02$ , $\beta_2 = 0.02 \pm 0.02$, $\gamma_2 = 0.59\pm 0.05$. These results are also listed in Table \ref{tab:table1}.  The error bars are deduced from the statistic uncertainties of the SFs extracted in the present work. According to the experimental as well as the theoretical definition of the intensity ($I$) ~\cite{Barker2009}, which is the SF divided by the adopted sum rule and hence sums up to $100\%$, and by using the expression of Ref.\,\cite{Kay-2013}, we have $I = SF_{exp}/[F_q\ast(2j+1)]$.  Based on the presently determined SFs (0.20 or 0.41) and intensities (0.19 or 0.39, respectively), the quenching factor $F_q$ can easily be deduced to be 0.53 for the $s$-wave ($j = 1/2$) components in the low-lying $0^+$ states of $^{12}$Be, fairly within the range of the nominal values \cite{Kay-2013}.

\begin{figure}
  \includegraphics[width=0.40\textwidth]{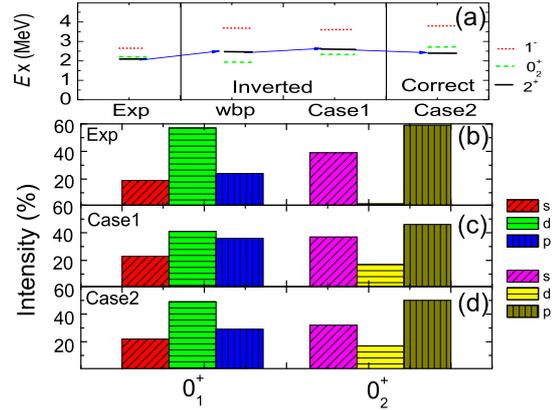}\\
  \caption{(a) Comparison of the level schemes of the low-lying states in $^{12}$Be between the experimental data and the shell
   model calculations with traditional wbp  \cite{Kanungo} or YSOX Hamiltonian. (b) The individual $s$-, $p$- and $d$-wave
    intensities for the 0$_1^+$ and 0$_2^+$ states deduced from experiments.  (c) Shell model calculations with YSOX
    interaction (Case1). (d) Same as (c) but with a decrease of 0.5 MeV for the $d$-orbit (Case2).}\label{occypancy}
\end{figure}

We have applied the shell model calculations, with the latest YSOX interaction \cite{Yuan,Yuan20162}, to reproduce the experimentally
 observed spectroscopic strengths. This approach works in a full $p$-$sd$ model space, including (0-3)$\hbar\omega$ excitations,
 and may give good descriptions of  the energy, electric quadrupole and spin properties of low-lying states in B, C, N, and O isotopes.
  The calculated individual $s$-, $d$-, and $p$-wave strengths for the first two 0$^+$ states in $^{12}$Be, denoted by Case1 in
  Fig.\ref{occypancy}(c), are compared to the experimental results shown in Fig.\ref{occypancy}(b). The calculated $s$-wave intensities
  for these two 0$^+$ states are in good agreement with the experimental ones, whereas the calculated $p$-wave intensity for the
   0$_1^+$(0$_2^+$) state is slightly larger (smaller) than the experimental value \cite{Meharchand}. This deviation in $p$-wave
   is opposite to the $d$-wave components. Despite a generally good description of intensities by the Case1 calculation,
   it does not give the correct level order of the low-lying excited states as demonstrated in Fig.\ref{occypancy}(a), neither does
    with the WBP interaction \cite{Kanungo}. A decrease of 0.5 MeV for the $d$-orbit in the calculation would lead to the restoration
     of the level order (a relative decrease of the $2^+$ state), and also a better reproduction of the  $p$-wave intensities, as
      displayed by Case2 in Fig.\ref{occypancy}(d). Case2 parametrization allows also a good description of the ground and low-lying
      excited states in $^{11}$Be. The meaning of this shift for $d$-orbit needs to be understood by further theoretical investigations.

\section{Summary}
 In summary, a new measurement of the $^{11}$Be($d$,$p$)$^{12}$Be transfer reaction was performed with a $^{11}$Be beam
 at 26.9$A$ MeV. Special measures were taken in determining the deuteron target thickness and in separating the $0_2^+$ isomeric
 state from the mixed excitation-energy peak. Elastic scattering of $^{11}$Be + $p$ was simultaneously measured to estimate the
 hydrogen contamination in the (CD$_2$)$_n$ target and to obtain the reliable OP to be used in the analysis of the transfer
  reaction. FR-ADWA calculations were employed to extract the SFs for the low-lying states in $^{12}$Be.  The ratio between
  the SFs of the two low-lying $0^+$ states, together with the previously reported results for the $p$-wave components, was
  used to deduce the single-particle component intensities in the two bound 0$^+$ states of $^{12}$Be, which are to be compared
   directly to the theoretical predictions. The results show a clear $d$-wave predominance in the g.s. of $^{12}$Be, which is
   dramatically different from the g.s. of $^{11}$Be dominated by a intruding $s$-wave. This exotic intruding phenomenon was also observed in a latest  $^{12}$Be($p$, $pn$) knockout reaction experiment \cite{LE-PLB-2017}. The present results are also compatible with those obtained from the previous transfer reaction measurements, considering the reported uncertainties. This work demonstrates
    the importance of measuring the individual SFs in the low-lying states in order to fix the configuration-mixing mechanism.

\section*{Acknowledges}
\label{acknowledgments}
We gratefully acknowledge the staff of RCNP accelerator group for
providing the $^{13}$C primary beam and the staff of EN-course
for the assistance and the local support. This work is supported by the National Key R$\&$D Program of China (the High Precision Nuclear Physics Experiments), the National
 Natural Science Foundation of China (Nos.11775004, 11775013, 11775316, 11535004, 11375017, and 11405005).

\end{document}